\documentclass[journal,comsoc]{IEEEtran}
\usepackage[T1]{fontenc}
\usepackage{cite}

\ifCLASSINFOpdf
   \usepackage[pdftex]{graphicx}
\else
   \usepackage[dvips]{graphicx}
\fi

\usepackage{amsmath}
\usepackage[cmintegrals]{newtxmath}

\usepackage[numbers]{natbib}
\usepackage{booktabs}
\usepackage{multirow}
\usepackage{multicol}
\usepackage{algorithm}
\usepackage{amsmath}
\usepackage{algorithmic}

\mathchardef\mhyphen="2D

\begin{document}
%
\title{Validation of the Results of Cross-chain Smart Contract Based on Confirmation Method}

\author{Hong~Su
\IEEEcompsocitemizethanks{\IEEEcompsocthanksitem H. Su is with the School of Computer Science, Chengdu University of Information Technology, Chengdu, China, 610041. \\
 E-mail: suguest@126.com. \\
\protect\\
}
\thanks{}}

\markboth{Journal of \LaTeX\ Class Files,~Vol.~14, No.~8, August~2015}%
{Shell \MakeLowercase{\textit{et al.}}: Bare Demo of IEEEtran.cls for IEEE Communications Society Journals}
%

\maketitle

\begin{abstract}
  Smart contracts are widely utilized in cross-chain interactions, where their results are transmitted from one blockchain (the producer blockchain) to another (the consumer blockchain). Unfortunately, the consumer blockchain often accepts these results without executing the smart contracts for validation, posing potential security risks. To address this, we propose a method for validating cross-chain smart contract results. Our approach emphasizes consumer blockchain execution of cross-chain smart contracts of producer blockchain, allowing comparison of results with the transmitted ones to detect potential discrepancies and ensure data integrity during cross-chain data dissemination. Additionally, we introduce the confirmation with proof method, which involves incorporating the chain of blocks and relevant cross-chain smart contract data from the producer blockchain into the consumer blockchain as evidence (or proof), establishing a unified and secure perspective of cross-chain smart contract results. Our verification results highlight the feasibility of cross-chain validation at the smart contract level.
\end{abstract}


%
\IEEEpeerreviewmaketitle

\section{Introduction}
Cross-chain technologies have become increasingly important in blockchain ecosystems, enabling seamless communication and data exchange across different blockchain networks \cite{ou2022overview}. They have been widely used in various industries, including financial services \cite{moncada2021next}, supply chain management \cite{peng2022research}, and healthcare \cite{wei2023study}. The cross-chain capabilities have helped improve information and value transfer, and have made digital transactions more transparent, secure, and efficient.

\begin{figure}[htb]
    \includegraphics[width=3.5in]{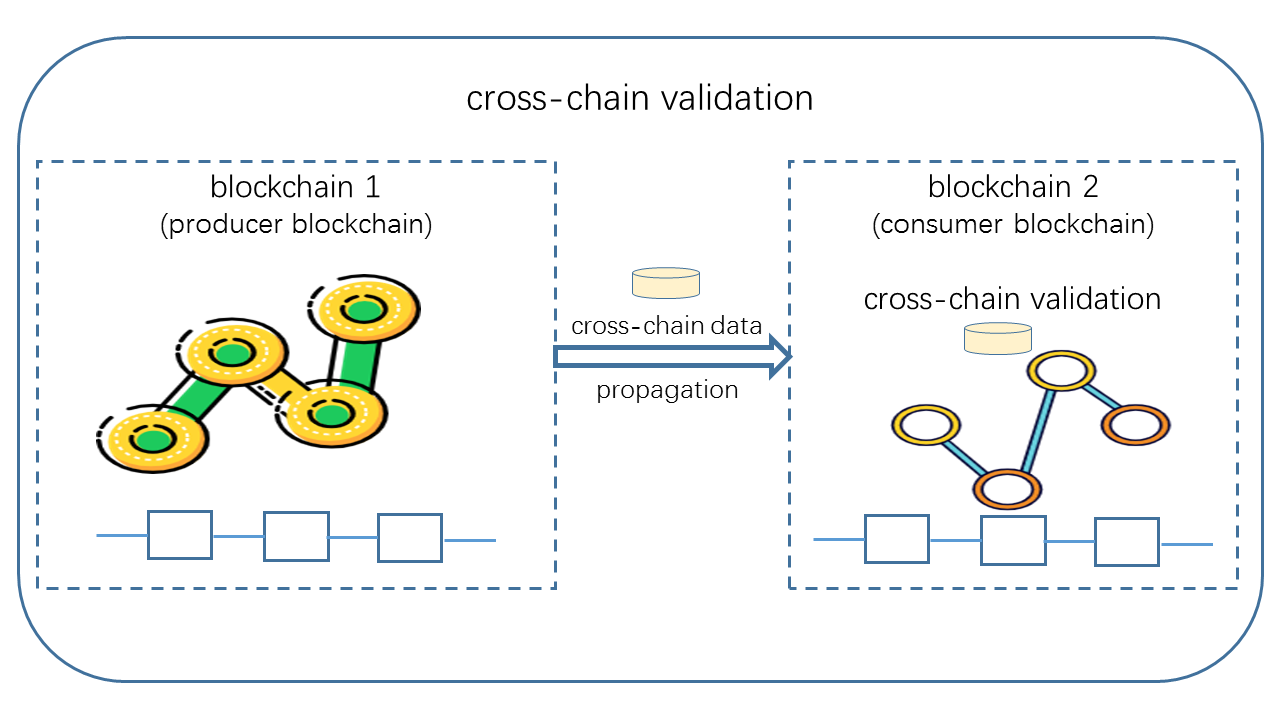}
    \caption{Cross-chain validation}
    \label{crosschain_verification}
\end{figure}

Cross-chain interaction refers to a fundamental concept in which the state or actions of one blockchain (referred as the producer blockchain in this paper) depend on the state or actions of another blockchain (referred to as the consumer blockchain in this paper). It is within this dynamic interaction that the potential for security challenges and risks arises. Ensuring the accuracy and trustworthiness of cross-chain data and the results of smart contracts is paramount, as any validation failure could lead to erroneous state changes in the consumer blockchain. Therefore, the validation of cross-chain data and smart contract results is of utmost importance in cross-chain scenarios. Figure \ref{crosschain_verification} shows one example of cross-chain validation.

Validating cross-chain smart contracts is a crucial prerequisite step in cross-chain interactions. Unlike validating simple cross-chain data, validating the results of smart contracts is far more complex. Smart contracts are intricate, self-executing agreements with predefined conditions and consequences, making it necessary to run the smart contract to verify its correctness. 

There are two challenging issues during the cross-chain smart contract validation:
(1) Cross-chain Validation method: Cross-chain validation is a critical process that typically involves executing smart contracts to ensure their integrity. While it's important for this process to execute smart contracts, it shouldn't disrupt the functioning of smart contracts within the consumer blockchain. Additionally, the execution of cross-chain smart contracts can consume resources within the consumer blockchain. In this context, the paper investigates cross-chain validation methods within separate running environments, taking into account resource utilization, to suit the requirements of various blockchain ecosystems.
(2) State Unification: Blockchain forks, which occur when a blockchain splits into multiple potentially conflicting chains of blocks, present a significant challenge during the validation process. In such scenarios, ensuring a unified state of the producer blockchain within the consumer blockchain becomes imperative. This paper introduces an innovative approach to confirm the chain of block information originating from the producer blockchain into the consumer blockchain. This confirmation process helps to unify the consumer blockchain's perspective of the producer blockchain.

The contributions of this paper are summarized as follows:

(1) Validation methods of cross-chain smart contract: This paper introduces a method for validating cross-chain smart contracts. Prior research has predominantly focused on validating blockchain data, with limited attention given to validating the results of smart contracts. However, the validation of cross-chain smart contract is a crucial aspect of ensuring the security and reliability of cross-chain interactions.

(2) Resource-efficient validation methods: To address concerns related to resource allocation, the validation methods consider the resources occupation. First, we propose the metrics which measure the resources occupied in the cross-chain smart contract validation. Second, we propose validation methods that consider different resource utilization ratios. One of the proposed methods, referred to as collective validation, entails multiple nodes that trust each other utilizing a single node to perform cross-chain smart contracts. These nodes share the validation results among themselves, promoting a collaborative and resource-efficient validation process.

(3) Cross-chain confirmation with proof: The method of confirmation with proof involves the integration of critical data from the producer blockchain into the consumer blockchain. This is accomplished by incorporating the chain of blocks and associated data from the producer blockchain as evidence (or proof) within the consumer blockchain. By securely embedding this information, the consumer blockchain gains a comprehensive and secure perspective of the cross-chain smart contract results. 

The remaining sections of this paper are organized as follows. In Section \ref{sec_related_work}, we provide an overview of related work in the field of cross-chain.
Section \ref{sec_ccv_sc} describes the method of the validation of cross-chain smart contract.
In Section \ref{sec_cc_conf}, we describe the method of cross-chain confirmation with proof for cross-chain smart contract.
In Section \ref{sec_verification}, we present the verification results and conduct corresponding analyses.
Finally, Section \ref{sec_conclusion} concludes this paper with a summary of our contributions and potential future research directions.

\section{Related Work} \label{sec_related_work}
In this section, we present a comprehensive review of existing approaches to cross-chain validation, encompassing both cross-chain data propagation, which involves the process of data validation, and various cross-chain data validation methods.

\subsection{Cross-chain Data Propagation Methods}

\subsubsection{User-Initiated Cross-chain Data Transfer}
User-initiated cross-chain data transfer represents a fundamental approach to cross-chain interactions. In this paradigm, users actively participate in transferring data from one blockchain to another, frequently employing techniques such as hash locking \cite{mao2022survey} where users use a secret number (or its hash) set in one blockchain as data to unlock assets in another blockchain. This method empowers users with a high degree of control and flexibility over the data transfer process.

In this approach, users initiate and oversee the entire data transfer operation. Transaction validation primarily hinges on the user's discretion and is carried out manually. Users decide when and how to initiate data transfer transactions, ensuring that the data being moved from one blockchain to another aligns with their intended use cases.

However, it's crucial to acknowledge that this method has its limitations. While it allows for direct user involvement and control, it simultaneously introduces the potential for human errors. Users must meticulously manage the data transfer process, including validating the data's integrity and ensuring it corresponds accurately to the destination blockchain's requirements. This manual approach becomes increasingly challenging and error-prone in scenarios involving intricate cross-chain interactions or frequent data transfers \cite{su2021cross}.

For complex cross-chain applications requiring a high level of reliability and data accuracy, the user-initiated data transfer approach may be less suitable. As such, alternative methodologies, such as oracle services and direct cross-chain data propagation, have emerged to address these limitations and streamline cross-chain data transfer processes.

\subsubsection{ Oracle Services}
Oracle services have rapidly gained prominence as indispensable tools for facilitating cross-chain data transfer \cite{lu2023ccio}. These services function as intermediary computing servers, serving the critical role of automating the retrieval of cross-chain data and its seamless propagation to other blockchain networks.

The central premise of oracle services revolves around automation and efficiency. By leveraging oracle solutions, blockchain ecosystems can significantly streamline the process of acquiring data from external sources and ensuring its availability on the target blockchain. This automation significantly reduces the manual intervention required from users, enhancing the overall efficiency of cross-chain interactions.

One of the hallmark features of oracle-based solutions is their autonomous transaction validation capabilities \cite{hei2022practical}. In this model, transaction validation is predominantly entrusted to the oracle service itself. This approach further reduces user involvement, as the oracle service acts as a trusted intermediary responsible for validating the data's integrity and compliance with predefined criteria.

Although oracle services offer numerous advantages in terms of automation and data reliability, they may introduce concerns related to centralization if not implemented with meticulous attention to decentralization mechanisms. Centralized oracle services could potentially become single points of failure or manipulation, undermining the decentralized ethos of blockchain technology.

To mitigate these centralization concerns, modern oracle solutions often incorporate decentralization features \cite{gigli2023decentralized}, such as multiple oracle nodes, data aggregation, and cryptographic validation. These mechanisms aim to preserve the integrity and security of cross-chain data transfer while maintaining a level of automation that rivals user-initiated approaches.

In summary, oracle services have emerged as a pivotal enabler of cross-chain data transfer, offering automation, reliability, and reduced user intervention. Nevertheless, their successful deployment hinges on striking a delicate balance between automation and decentralization to preserve the core principles of blockchain technology.

\subsubsection{Direct Cross-Chain Data Propagation}
Direct cross-chain data propagation \cite{su2021strongly} has been facilitated by specialized components such as blockchain routers \cite{wang2017blockchain}, ushering in a new era of seamless data exchange and interaction between disparate blockchain networks. These innovative components serve as the backbone for enabling nodes from distinct blockchains to communicate directly and share data efficiently.

At the heart of this approach lies the concept of decentralized autonomy. Rather than relying on centralized intermediaries or user-initiated actions, direct cross-chain data propagation empowers the participating nodes themselves. These nodes, representing distinct blockchains, take an active role in the verification and validation of incoming data \cite{su2021cross}.

The transaction validation process in direct cross-chain data propagation is notably distributed across these participating nodes. Each node is tasked with verifying the data relevant to its respective blockchain, ensuring its consistency, accuracy, and adherence to predefined rules and protocols. This collaborative validation approach effectively combines the benefits of automation with the principles of decentralization.

One of the key strengths of this approach is its ability to strike a harmonious balance between automation and decentralization. By leveraging the collective computational power and trustworthiness of nodes across different blockchains, direct cross-chain data propagation offers a robust and reliable means of data transfer and validation. This makes it an ideal choice for a wide spectrum of cross-chain applications, from simple asset transfers to more complex, multi-chain smart contract executions.

While direct cross-chain data propagation exhibits significant promise, it is not without its challenges. Coordination among diverse blockchain networks and ensuring seamless interoperability can be technically intricate. Furthermore, robust security mechanisms must be in place to safeguard data during transmission and validate it effectively. Nevertheless, as this approach matures and becomes more widely adopted, it holds the potential to revolutionize how cross-chain interactions are conducted, fostering a new era of blockchain connectivity and collaboration.

In summary, direct cross-chain data propagation, facilitated by components like blockchain routers, exemplifies the ideal marriage of automation and decentralization. It empowers nodes from distinct blockchains to work together in validating and propagating data, offering a versatile solution for cross-chain applications while maintaining the fundamental principles of blockchain technology.

\subsection{Cross-chain Data Validation}
\subsubsection{Atomic Swaps} Atomic swaps allow different blockchains to swap assets without any intermediary \cite{manevich2022cross}. These swaps are atomic, meaning that they either occur completely or not at all. 
The advantage of atomic swaps is that they provide trustless asset exchange between different blockchains without the need for centralized exchanges or third parties. However, atomic swaps may require significant resources and time to set up and execute. 
Data validation in atomic swaps occurs through the execution of pre-agreed rules and conditions agreed upon by both parties. These rules are typically encoded in smart contracts on both blockchains involved in the swap. The smart contracts ensure that the swap conditions are met before assets are released to the other party. Atomic swaps typically require both parties to cryptographically prove their holdings of assets to be swapped.

\subsubsection{Hash Time-Locked Contracts (HTLC)} HTLCs allow two parties to exchange value across blockchains without trusting each other \cite{monika2022hash}. HTLCs typically involve a hash commitment scheme where both parties commit to a random number or hash value before exchanging the assets. If the hashes match at a later time, the assets are released to the other party. HTLCs are commonly used in atomic swaps and provide trustless asset exchange between different blockchains. However, HTLCs may be vulnerable to double spending attacks if not implemented properly.
Data validation in HTLCs occurs through hash commitment schemes where both parties commit to random numbers or hashes before exchanging assets. The hashes serve as commitments to the transaction details, ensuring that both parties have agreed on the terms of the exchange. If both parties successfully match their hashes at a later time, the assets are released to the other party. HTLCs typically require both parties to cryptographically prove their holdings of assets to be swapped.

\subsubsection{Lightning Networks} Lightning networks are layer-2 solutions for Bitcoin and other blockchains that allow for fast and low-cost transactions \cite{wu2022increasing} \cite{ren2018optimal}. They operate as payment channels between users on the blockchain network and can be used for cross-chain data validation. Lightning networks provide privacy and scalability advantages over traditional blockchain transactions and allow for faster and cheaper transactions. However, they may be vulnerable to routing attacks if not implemented properly.
Data validation in lightning networks occurs through multiple stages of transaction confirmation and validation between network nodes. Each transaction on the lightning network is typically confirmed multiple times by different nodes, forming a path of trustless routing nodes called "channels" that route transactions between users on the blockchain network. These channels maintain transactional records that are cryptographically encrypted and synchronized across participating nodes on the network, ensuring trustless operation and data validation.

\subsubsection{Polkadot} Polkadot is a new blockchain protocol that connects different blockchains together using relays \cite{zhang2023cross}. It allows cross-chain data validation and transfer of assets between different blockchains. Polkadot's architecture includes a relay chain (or "canister") that anchors blockchains together, enabling interoperability between different chains. Relay chains validate and authenticate transactions between different chains, ensuring security and trustless operation. Polkadot also provides future compatibility for existing blockchain technologies such as Ethereum and Bitcoin, allowing them to interoperate with Polkadot's relay chain.
Data validation in Polkadot occurs through relay chains that anchor blockchains together and validate transactions between different chains \cite{wang2020electricity}. Relay chains maintain transactional records and cryptographically encrypted records of each transaction to ensure trustless operation across multiple chains. Additionally, Polkadot's architecture includes validation nodes called "canister nodes" that validate transactions and cryptographically seal them into the blockchain, ensuring data integrity and validation on the relay chain level.

\subsubsection{Cosmos} Cosmos is also a blockchain network that connects different blockchains together using hubs and zones \cite{qasse2019inter}. It provides cross-chain data validation and atomic swaps between different blockchains. Cosmos' architecture includes hubs (or "hub chains"), which serve as central coordinators for multiple zones (or "zone chains") connected to them \cite{yang2018cvem}. Hub chains validate transactions and coordinate zone chains to ensure security and trustless operation across multiple chains. Atomic swaps can be performed between different zones connected to the same hub or between zones connected to different hubs within the Cosmos network.
Data validation in Cosmos occurs through hub chains that serve as central coordinators for multiple zones connected to them. Hub chains validate transactions and coordinate zone chains to ensure trustless operation across multiple chains connected to them. Atomic swaps can be performed between different zones connected to the same hub or between zones connected to different hubs within the Cosmos network, with data validation occurring through pre-agreed rules encoded in smart contracts on hub chains and zone chains involved in

\subsection{Gap in Existing Approaches}
While the aforementioned approaches effectively address various aspects of cross-chain data validation, they often lack emphasis on the validation of smart contract. As a consequence, potential issues pertaining to the accuracy and reliability of cross-chain smart contract outcomes remain unaddressed. In this paper, we direct our focus towards this critical aspect, proposing methods to enhance the security and dependability of smart contract executions in cross-chain scenarios.

\section{Cross-chain Validation of Results of Smart Contract} \label{sec_ccv_sc}
Cross-chain validation involves the verification of data or smart contracts from other blockchains, which can be categorized into two corresponding aspects: data validation and smart contract validation.
Data validation entails the verification of cross-chain data originating from the producer blockchain, encompassing cross-chain-related transactions and blocks. This process confirms the occurrence of specific states within the producer blockchain.

The characteristics of cross-chain data validation is that there is no requirement of the execution of a smart contract. For example, a transaction in the chain of blocks of the producer blockchain, it can be verified by whether it is sent by the corresponding sender and whether the sender has enough balance or not. Contrarily, the validation of a smart contract includes the step to run the smart contract.

Smart contract validation is the process of ensuring that the results of a smart contract's execution are accurate and valid. The aim is not to run the smart contract, but to check whether the corresponding results are correct or not. In the context of cross-chain interactions, it is crucial to verify the results of cross-chain smart contracts  as the results will cause corresponding actions in consumer blockchain.

There are two different ways for the validation of smart contract, smart contract level validation (SCL validation) and block level validation (BL validation).
In the smart contract level validation, it is to run the smart contract and get the corresponding results to compare whether it is the same as the results in the block.

In the block level validation, it is just to check whether the block is correct (such as whether it is in the longest chain of blocks) and trust the block without running the smart contract. A block in the longest chain is deemed valid, and its associated data, including transactions and smart contract results, is considered valid as well. However, this straightforward validation method is not always applicable in cross-chain scenarios, as there is no equivalent mechanism in place during cross-chain propagation.

The reason to use the SCL validation is for the following three reasons. (1) In certain situations, results are calculated based on multiple states recorded in the blockchain, rather than through direct access. If the smart contract from the producer blockchain is not executed, a similar logic must be recreated on the consumer blockchain to derive the results from the related states. (2) In some blockchains, only the hash of results of smart contract are written to the blockchain, the consumer blockchain cannot use the hash for cross-chain interact directly. (3) When a miner seals in wrong results in the longest blockchain, it will take time for other nodes to correct the results, thus if the corresponding smart contract is not run directly, it will take more time to find the wrong results.

However, there is currently a lack of research focused on the validation of the results of cross-chain smart contracts. This type of validation is crucial since the majority of existing blockchains employ smart contracts for implementing digital contracts, including those designed for cross-chain scenarios, and the results of these smart contracts hold importance in cross-chain interactions.

\subsection{Validation in Smart Contract Level}
The smart contract level validation is that the nodes in the producer blockchain run the smart contract of the producer blockchain to verify the results to avoid errors or cheating during cross-chain propagation.
However, the resources required by smart contracts of consumer and producer blockchains may be conflict, they should run in isolated environments. Meanwhile, there are interacts between them, there should be cross environments methods to let nodes of consumer blockchain get the results of execution.

\subsubsection{Running Cross-chain Smart Contract in Separate Environment}
Running in a separate environment implies that cross-chain smart contracts from the producer blockchain are executed in an isolated setting. This distinct environment may leverage various technologies, including containers, virtual machines, or isolated runtimes. It's essential to clarify that the term "cross-chain smart contract" in this context specifically refers to smart contracts from producer blockchains, distinguishing them from smart contracts that are native to the local blockchain but used for cross-chain interactions.

The process of instantiating and running the smart contract involves the actual execution of the contract's logic, using the provided byte code and data to bring it to life within the blockchain environment.

(1) Cross-chain Instantiation: The consumer blockchain synchronizes the transaction that carries the instantiation parameters (or deployment transactions), including the binary code of the smart contract, instantiation parameters, and related data. Cross-chain instantiation encompasses the creation of an instance of the cross-chain smart contract within a dedicated runtime environment on the consumer blockchain. During instantiation, the contract's byte code and any accompanying data are utilized to establish the initial state of the contract.

(2) Invoking the Contract for Validation: After the smart contract is instantiated, it is activated by executing its methods or functions using the corresponding parameters. These parameters are conveyed via transactions, and as part of the cross-chain validation process, the corresponding invocation transactions must also be synchronized from the producer blockchain. It's important to note that this invocation is primarily for validation purposes, and all transactions have been previously recorded in the producer blockchain rather than being initiated interactively by the user.


(3) Result Generation and Validation: As the execution progresses, the smart contract generates results that represent the outcomes of the contract's operations. Different from the producer blockchain, these results are used to verify the results of the producer blockchain instead of the state changes, transaction outputs, or data updates. The consumer blockchain will use the cross environment method provided by the environment of the producer blockchain to get the results, for example by using the HTTP query method.


\subsection{Resource Optimization for Cross-Chain Smart Contracts}

\subsubsection{Collective Validation of Producer Blockchain}
In the proposed model, it's worth highlighting that not every node in the consumer blockchain network needs to maintain a dedicated environment for running validation smart contracts within the producer blockchain. 
Maintaining separate producer environments for each consumer node can result in significant resource redundancy, including duplicated storage space, computational power, and memory.
Instead, several nodes can efficiently share a single producer environment. This approach is especially advantageous for nodes that share similar interests or benefits, such as some nodes within the same research laboratory or company, all managed by a single entity or person.

This optimization offers substantial resource savings, reducing the overall computational and storage resources required to execute and verify smart contracts in the consumer environment. The corresponding validation method is called collective validation. 

To illustrate the resource efficiency gained by collective validation, we can formulate the following equation:

Let:
- \(N\) be the total number of consumer nodes in the network.

- \(R_{\text{individual}}\) represent the resources required for an individual producer environment.

- \(R_{\text{shared}}\) denote the resources needed for a shared producer environment.

- \(S\) be the total resource savings achieved through sharing environments.

The resource savings (\(S\)) can be calculated as in \eqref{eq_resource_occupation}.

\begin{equation} \label{eq_resource_occupation}
    S = N \cdot (R_{individual} - R_{shared})
\end{equation}

This equation demonstrates that the total resource savings (\(S\)) scale linearly with the number of consumer nodes (\(N\)). By adopting a shared environment strategy, these savings can be significant, reducing the overall resource overhead in the network.

The collective validation approach optimizes the utilization of computational, storage, and memory resources within a cross-chain interaction framework. It promotes resource efficiency, cost savings, scalability, and network resilience while ensuring that the validation of smart contracts remains robust and secure.

\subsubsection{Utilizing Cross-Chain Embedded Smart Contracts for Resource Optimization}
In the smart contract validation, there are two main aspects we should consider for the consumer blockchain.
(1) Calculations resources occupation. From the above process, we can see that a smart contract is loaded into the memory, instantiation, and running on the nodes of the consumer blockchain. This process occupies corresponding resources, including as CPUs and memories. Together with the smart contract on the consumer blockchain, it will burden the nodes of consumer blockchain.

(2) Synchronization data burden. Meanwhile, there are transactions used to trigger the interface of a smart contracts, which carries the parameters to invoke the interface, and we call these transactions the invocation transactions. While these transactions may locate in different blocks, which is then required for the validation of the execution results of smart contracts. The more invocation transactions required, the more information that requires to be synchronization during the cross-chain validation.

Considering the above two burdens, we give to measurements for the cross-chain smart contract validation, occupation time, related transactions, which can described as a two-tuples $burdern$, as shown in \eqref{eq_def_burden_tuples}.

\begin{equation} \label{eq_def_burden_tuples}
burdern = (Time_{occupation}, transaction_{numbers}).
\end{equation}
, where $Time_{occupation}$ is the resource occupation time and $transaction_{numbers}$ is the transactions that are requires to verify the results of the smart contract.

We now discuss resources occupation time, $Time_{occupation}$ to measure the time that a smart contract occupies resources of blockchain nodes. $Time_{occupation}$ is shown in \eqref{eq_def_time_occupation}.

\begin{equation} \label{eq_def_time_occupation}
    Time_{occupation} = Time_{end} - Time_{begin}
\end{equation}
, where $Time_{begin}$ and $ime_{end}$ are the time that a smart contract is instantiated and the time that the smart contract is terminated separately.

If a smart contract can have less $Time_{occupation}$, it can reduce the resources' occupation of blockchain nodes. The following section will discuss one kind of this smart contract.

Meanwhile, research indicates that a significant proportion of smart contracts, around 75\%, remain uninvoked, with 80\% only being executed once. One approach to mitigate the resource consumption of these smart contracts is through the use of embedded smart contracts. Embedded smart contracts are introduced in the work by Suh et al. \cite{su2022embedding}, a method that integrates the smart contract within its first invocation transaction. In this section, we evaluate the advantages and potential resource savings associated with the use of embedded smart contracts for cross-chain interactions.

Embedded smart contracts are instantiated exclusively during invocation, providing an effective means to conserve resources. This conservation applies to both computational and transaction-related resources, which are vital considerations in cross-chain interactions.

(1). Resource Efficiency: One key benefit of embedded smart contracts is the efficient utilization of computational resources. As illustrated in equation \eqref{eq_cmp_time_occupation}, which outlines the time relationship, the embedded smart contract's code is initiated later than traditional deployments ($Time_{begin} < Time_{putting}$). Consequently, it imposes a reduced burden on the computational resources of blockchain nodes.

\begin{equation} \label{eq_cmp_time_occupation}
    Time_{begin} \approx Time_{invocation} < Time_{putting}
\end{equation}

(2). Transaction Optimization: The use of embedded smart contracts streamlines the validation process by reducing the number of cross-chain transactions required. Traditional deployment necessitates separate transactions for contract deployment and its initial invocation. With embedded smart contracts, these two steps are merged into a single transaction, effectively reducing the number of transactions required for validation.

Moreover, there exists a unique class of smart contracts known as "disposable smart contracts." These contracts possess a distinctive trait: they are designed to be invoked only once and are immediately terminated, precluding any subsequent invocations. This characteristic sets them apart from conventional long-lived smart contracts.
This unique usage pattern allows for a resource-saving strategy. When a smart contract can be identified as disposable, it significantly reduces the resources expended in cross-chain interactions, particularly with regard to both time and transactions.

Disposable smart contracts require minimal resource allocation due to their ephemeral nature. Unlike long-lived contracts, which maintain a presence on the blockchain, disposable smart contracts have a concise lifecycle. As seen in Equation \eqref{eq_cmp_time_disposable_occupation}, the smart contract's initiation (begin) closely aligns with its invocation and termination times ($Time_{begin} \approx Time_{invocation} \approx Time_{end}$). This near-simultaneity results in the efficient use of resources.
\begin{equation} \label{eq_cmp_time_disposable_occupation}
Time_{begin} \approx Time_{invocation} \approx Time_{end}
\end{equation}

\section{Cross-chain Confirmation with Proof for Results of Cross-chain Smart Contract} \label{sec_cc_conf} 
Following the validation of cross-chain smart contract results, there is a need to verify and put these results and their associated information into the consumer blockchain. It's important to note that cross-chain smart contract results are not inherently present in the consumer blockchain's blockchains. Cross-chain confirmation with proof for cross-chain smart contract has two characteristics: (a) to confirm the results of cross-chain smart contracts within the consumer blockchain (confirmation), and (b) to provide evidence that these results have also been verified and accepted within the producer blockchain (proof).

The requirement for confirmation arises to accomplish two primary objectives:

(1) Mitigating Deception: Confirmation entails providing substantial proof that the relevant events transpired on the producer blockchain. By confirming this information, consumer blockchain nodes enhance security and guard against deceptive actions on the producer blockchain. In scenarios where a consumer blockchain attempts to manipulate a block and later alleges that it has undergone re-branching, the validation process ensures the veracity of such claims.

(2) Ensuring a Unified View of the Producer Blockchain: Confirmation guarantees that the nodes of the consumer blockchain maintain a coherent and unified perspective of the producer blockchain. Both producer and consumer blockchains can experience re-branching events, leading to two significant benefits:
(2a) When a specific block in the producer blockchain disappears, without corresponding confirmation, the nodes responsible for mining the block that contained the vanishing cross-chain information may be unjustly accused of wrongdoing.
(2b) In cases where a block containing a specific cross-chain event from the consumer blockchain disappears, the importance of cross-chain confirmation and proof mechanisms becomes evident. When such a block vanishes, it raises the need for a unified view within the consumer blockchain regarding whether the particular event has indeed occurred on the producer blockchain. Confirmation with proof processes are vital to maintain this unified perspective and ensure the security and integrity of cross-chain interactions.


Now, we provide the definition of cross-chain confirmation with proof, which entails confirming essential information from the producer blockchain to the consumer blockchain. This confirmation includes supporting evidence that specific events have occurred on the producer blockchain, ultimately guaranteeing the precision of cross-chain interactions.

Cross-chain confirmation was initially introduced in the context of conditional transactions, as discussed in \cite{su2022cross}. However, this method primarily focuses on confirming individual transactions and does not encompass comprehensive block confirmation with supporting proof. 

\subsection{Prerequisites for Cross-chain Confirmation}
Several prerequisites are essential for enabling cross-chain confirmation.

\subsubsection{Longest Chain Rule} The primary prerequisite is the adoption of the longest blockchain rule \cite{su2022cross} \cite{nakamoto2008bitcoin}. This rule dictates that, when confronted with two conflicting chains of blocks, the longest chain is recognized as the main chain. In other words, the chain where the data is widely accepted. Since the longest chain typically boasts the highest number of mining nodes and is the most challenging to disrupt, it garners widespread acceptance and is regarded as the only valid blockchain.

\subsubsection{Difficulty in Forging Blocks} It is crucial to make it challenging to forge a block, imposing certain constraints that deter illicit actions by hackers \cite{}. In Proof of Work (PoW), miners must continuously compute to discover a nonce that results in the block's hash meeting specific criteria. In Proof of Stake (PoS), an account's likelihood of mining a block is proportional to the assets it holds. This makes it challenging for hackers to forge the signatures of target accounts.

\subsubsection{Confirmation Completion Judgment} To declare that information is confirmed, there must be a sufficient number of successive blocks verifying the previous ones. The addition of a new block to an older one makes it increasingly difficult to falsify the older block, as demonstrated in \cite{nakamoto2008bitcoin}. When multiple blocks follow a specific block, it becomes proportionately more challenging to falsify all subsequent blocks.

\subsection{Requirements for Cross-chain Confirmation} \label{sec_require_cc}
In order to confirm the results of cross-chain smart contracts, the following are two requirements for cross-chain confirmation.

(1) The validation process requires the entire relationship of chain of blocks from the producer blockchain. This approach ensures an equivalent level of security as the original blockchain (the producer blockchain). 

This requirement is crucial to prevent the mining nodes of the consumer blockchain from falsifying data related to the producer blockchain. The chain of blocks constitutes successive evidence chain for every event within the producer blockchain. Any omission of blocks would render it impossible to establish the integrity of subsequent blocks, starting from the genesis block.

To illustrate this, let's assume there are $n$ blocks in the producer blockchain. If we decide to skip the synchronization of a block, indexed as $m$ (where $1 < m < n$), the synchronized chain's level of difficulty is equivalent to a chain comprising $n-m$ or $m-1$ blocks, rather than the complete $n$ blocks. Consequently, while the blocks from 1 (the genesis block) to $m-1$ are synchronized, they do not contribute to the verification of the blocks from $m$ to $n$.


(2) The information necessary for cross-chain confirmation should be minimized, focusing exclusively on synchronizing the data required for the relationships among blocks and validating the results of cross-chain smart contracts.

Regarding the chain of blocks, complete synchronization of all block details is unnecessary. Only the block headers need to be synchronized, as they contain sufficient information to verify the block's authenticity and difficulty. Therefore, in this paper, we specifically refer to block headers rather than the entire block when discussing the chain of blocks.

For transactions, synchronization is restricted to those transactions essential for cross-chain validation. This includes deployment transactions, invocation transactions, termination transactions, and similar transactions related to cross-chain smart contracts.
When synchronizing a cross-chain transaction, the associated Merkel tree, as described in \cite{nakamoto2008bitcoin} (or other relevant sources), must also be included to establish the transaction's validity.

\subsection{Cross-Chain Confirmation with Proof}
Cross-chain confirmation with proof is to mine data from the producer blockchain into the consumer blockchain to the consumer blockchain. This process comprises two main phases: the data synchronization process and the data mining process.

The data synchronization process, as discussed in \cite{su2021strongly}, entails synchronizing data from the producer blockchain to the consumer blockchain. We won't delve into further details in this paper, as it has been addressed in \cite{su2021strongly}.

Following data synchronization, miners must incorporate this data into the consumer blockchain. We refer to this step as the cross-chain mining process. Essentially, this process involves mining the data, including the necessary blocks and transactions, from the producer blockchain into the consumer blockchain.

Block headers are periodically synchronized and mined into the consumer blockchain. It's crucial that newly confirmed blocks from the producer blockchain form a continuous chain starting from the last confirmed chain of blocks on the producer blockchain. This means that if the last confirmed block was $bl_k$, the next block should be $bl_{k+1}$, and it must have a hash link in the blockchain header to the previous block, $bl_k$.

However, since the primary objective is to validate the producer data, there's no necessity to mine data from the producer blockchain into the consumer blockchain on a block-by-block basis. A more efficient approach is to mine data at regular intervals, for example, every $n$ blocks (where $n$ is a specific number), or when a block contains cross-chain data. A collection of blocks from the producer blockchain that are mined at once in the consumer blockchain is referred to as a \textbf{cross-chain block segment}, and $n$ represents the length of a cross-chain block segment. This strategy helps mitigate the challenges associated with frequent re-branching in the producer blockchain.

Let's illustrate this with a brief example. Imagine there are $n$ blocks of producer blockchain, denoted as $bl_{m+1}$, $bl_{m+2}$, ..., $bl_{m+n}$, which have not yet been mined in the consumer blockchain. When later these $n$ blocks are mined, consider the first block, $bl_{m+1}$. It now has $n-1$ consecutive blocks following it, which significantly raises the difficulty for any attempt to re-branch.

For instance, if we set $n$ to 3, the first block ($bl_{m+1}$) will have 2 successive blocks ($bl_{m+2}$, $bl_{m+3}$) linked to it, making it far more resistant to re-branching than a block with no successive blocks.

While rebranching is only one of the impact factors, selecting an appropriate value for $n$ should adhere to the following criteria.

(R1)The total size of the data from the producer blockchain, represented by $n$ pieces, should be smaller than the maximum block size of the consumer blockchain to ensure it can fit within a single block. This can be expressed as:

    \begin{equation}
        n * sizeof(block header) < sizeof (block)
    \end{equation}

(R2) If condition (R1) can be met, then $n$ should be a number that is either close to or greater than the number of successive blocks required to confirm a block. This is because the first blocks in this block segment have the enough low probability of being rebranched. This can be framed as:

    \begin{equation}
        {(p_{fake}^{avg})} ^ n  < \delta
    \end{equation}
    , where $p_{fake}^{avg}$ represents the average probability of a block being fake.

(R3) On the other hand, $n$ should not be excessively large, as indicated in \eqref{not_too_big}. When $n$ is too large, there's a risk that other nodes may have ample time to fabricate a cheating block. This is particularly relevant as we allow the mining of one block if it contains cross-chain smart contracts.

    \begin{equation} \label{not_too_big}
        n * time_{block}^{avg}  < \beta
    \end{equation}
    , where $time_{block}^{avg}$ denotes the average time for a block to be produced in the producer blockchain.





\subsection{Block Data Structure for Cross-chain Confirmation}
\begin{figure*}
  \centering
  \includegraphics[width=6in]{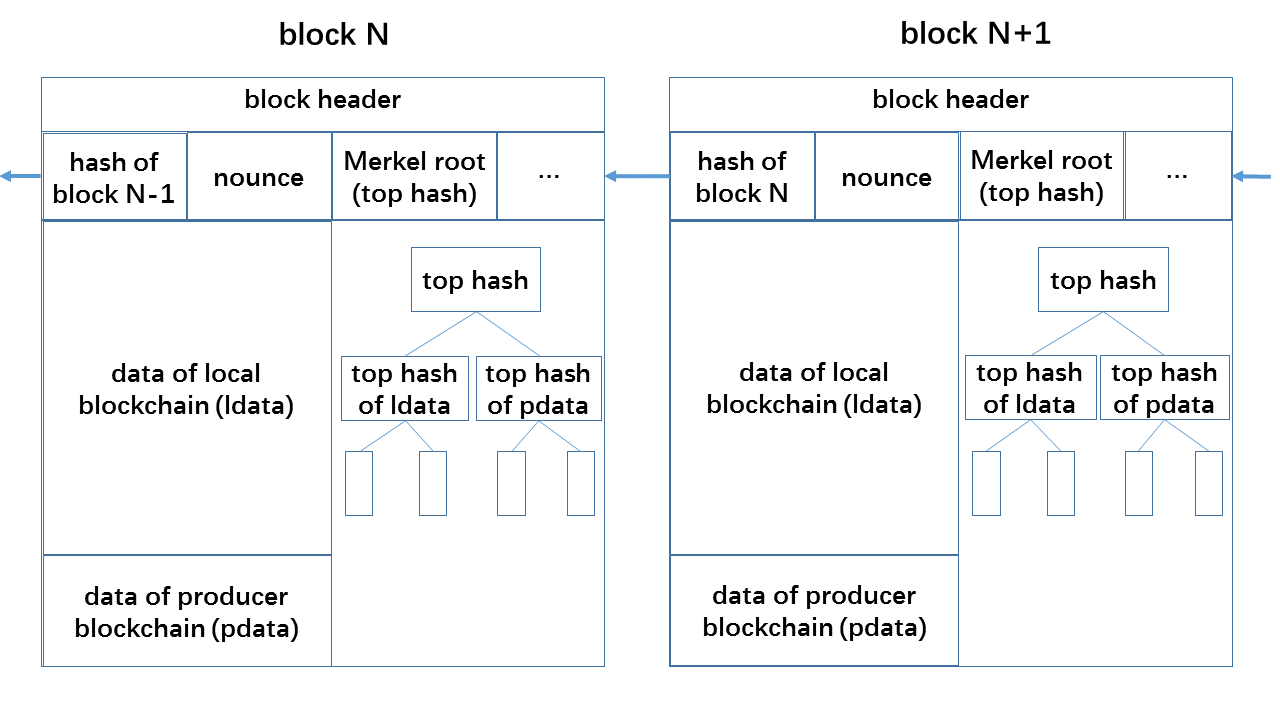}
  \caption{The new block data structure with data of consumer blockchain}
  \label{block_strcuture}
\end{figure*}

To facilitate the cross-chain confirmation process, the block data structure of the consumer blockchain will be adjusted to include data used for validation from the producer blockchain: the chain of blocks of the producer blockchain and the necessary data to validate the results of cross-chain smart contracts. A depiction of the block structure is shown in Figure \ref{block_strcuture}.

To support the confirmation with proof, the Merkle tree is specially designed, with the root having two subtrees. One subtree is for internal data (such as the hash of the consumer blockchain), while the other is for the producer blockchain. This design prevents the chain of blocks from reverting to the producer blockchain, which could cause confusion.

\section{Security Analysis}
In the proposed model, a fundamental security enhancement is achieved through the validation of cross-chain smart contract results and confirmation with proof. We attempt to analyze the corresponding impact from the corresponding aspects.

\subsection{Validation of Cross-Chain Smart Contract Results}
\subsubsection{Mitigating Cheating}
By independently verifying the results of cross-chain smart contracts, the consumer blockchain can ensure that it is not reliant on potentially untrustworthy or erroneous data from the producer blockchain. This is crucial for maintaining the integrity of cross-chain interactions. If the results were not fully verified, dishonest nodes on either the producer or consumer blockchain could potentially manipulate or cheat the system, leading to incorrect outcomes and security vulnerabilities.

One crucial motivation for cross-chain smart contract validation is the absence of an assurance during cross-chain synchronization, despite the presence of validation processes within each blockchain. Consequently, with cross-chain smart contract validation, any discrepancies can be detected, and the validation process can prevent incorrect or tampered data from being incorporated into the consumer blockchain.

We can illustrate this briefly. Let \(Data_{producer}\) represent data from the producer blockchain. \(Con_{producer}\) represents the consensus algorithm of the producer blockchain. \(S\) represents the synchronization process, which transfers data from the producer to the consumer blockchain.

The validation process \(V\) ensures the trustworthiness of data from the producer blockchain, as shown in Equation \eqref{eq_propagation_process}.

\begin{equation} \label{eq_propagation_process}
  V(S(Data_{producer}), Con_{producer}) \rightarrow \text{Verified}
\end{equation}

Here, \(V\) ensures that the data received via synchronization (\(S(Data_{producer})\)) matches the consensus algorithm on the producer blockchain (\(Con_{producer}\)).

The process illustrated by Equation \eqref{eq_propagation_process} reinforces the critical role of validation in enhancing security and mitigating cheating risks in cross-chain interactions.



\subsubsection{Data Integrity}

Verifying the results of cross-chain smart contracts ensures data integrity within the consumer blockchain. When cross-chain smart contract results are verified, it provides an additional layer of assurance that the data stored in the consumer blockchain is accurate and trustworthy. This validation mechanism acts as a safeguard against potential data tampering or inaccuracies that could occur during the cross-chain data transfer process. By verifying results, the consumer blockchain can maintain a high level of data integrity, enhancing the overall reliability and security of the blockchain network. This is especially important for applications where the accuracy of data is critical, such as financial transactions or supply chain management. Any compromise in data integrity can lead to financial losses or other adverse consequences.

\subsection{Maintaining or Enhancing the Security of Producer Blockchain Data in the Consumer Blockchain}
The model aims to ensure that the security of data from the producer blockchain is not compromised in the consumer blockchain. 

To confirm data from the producer blockchain in the consumer blockchain, nodes on the consumer blockchain must provide successive blocks that match the corresponding difficulties. Attempting to introduce fraudulent data would require significant computational resources and would be economically unfeasible, as illustrated in Equation \eqref{eq_cost_relationship}. This is due to the mining process, which involves solving cryptographic puzzles that requires substantial computational resources.

\begin{equation} \label{eq_cost_relationship}
  cost_{fraud} > gain_{fraud}
\end{equation}
, where \(cost_{fraud}\) means the cost to make a fraud data of the producer blockchain, and \(gain_{fraud}\) means the corresponding benefits.
\\

In summary, the model's approach to fully verifying cross-chain smart contract results and maintaining or enhancing the security of producer blockchain data in the consumer blockchain significantly bolsters the security and trustworthiness of cross-chain interactions. It mitigates cheating, error handling issues, and ensures data integrity, making it suitable for a wide range of applications where data accuracy and security are paramount.

\section{Verification} \label{sec_verification}
In this section, we aim to verify the feasibility of the proposed solution on two aspects. First,  we assess the resources utilized in the validation of cross-chain smart contracts and the implemented methodologies to conserve resources. This verification aims to demonstrate the efficiency of resource usage and the viability of cross-chain smart contract validation. Second, we evaluate the impact of cross-chain segment block lengths on cross-chain smart contract validation, as it is a crucial factor affecting the confirmation process.

\subsection{Resource Consumption of Cross-chain Smart Contract Validation}
The resource consumption of cross-chain smart contract validation was assessed in this section to evaluate its feasibility from the perspective of resource utilization. The verification focused on CPU and memory usage. 

We conducted a comparison of resource utilization between two blockchains, denoted as blockchain1 and blockchain2. Both blockchains utilize the Proof of Work (PoW) consensus algorithm and synchronize their transactions and blocks. To assess resource consumption, we configured nodes within blockchain1 for distinct purposes. One node was dedicated solely to cross-chain smart contract validation without engaging in mining operations, labeled as \textbf{CCValidationWithoutMining}. Another node was exclusively assigned to mining activities without participating in cross-chain smart contract validation, identified as \textbf{MiningWithoutCCValidation}. The remaining nodes were tasked with both mining and cross-chain validation, classified as \textbf{MiningAndCCValidation}.

The cross-chain smart contract is designed to perform the following operations, aiming to simulate tasks that take a certain amount of time.
(1) Generate a random number between 100 and 100 * 100 * 100, denoted as $scLoopTimes$.
(2) Loop from 1 to $scLoopTimes$ to generate random numbers and accumulate their sum.
(3) Output the total sum after the loop is completed.

A periodic task triggers transactions to call the smart contract every 3 seconds, aiming to generate multiple trigger transactions within one mining period, which typically takes around 10 seconds. Another task utilizes the 'top' command in Linux to record CPU and memory utilization every 5 seconds, capturing resource usage during mining.

\begin{figure}
    \includegraphics[width=3.5in]{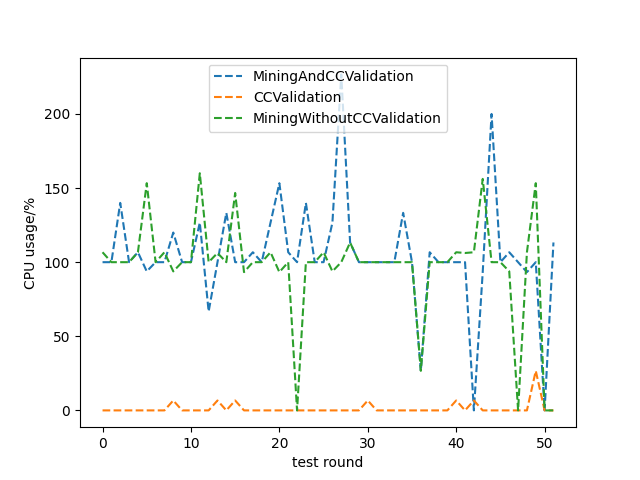}
    \caption{The CPU occupation comparison.}
    \label{resourceOccupationCPU}
  \end{figure}

\begin{figure}
    \includegraphics[width=3.5in]{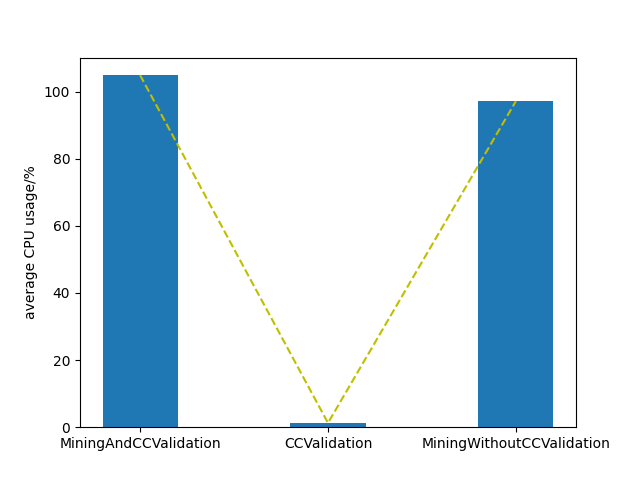}
    \caption{The average CPU occupation comparison.}
    \label{resourceOccupationCPUAverage}
  \end{figure}

\begin{figure}
  \includegraphics[width=3.5in]{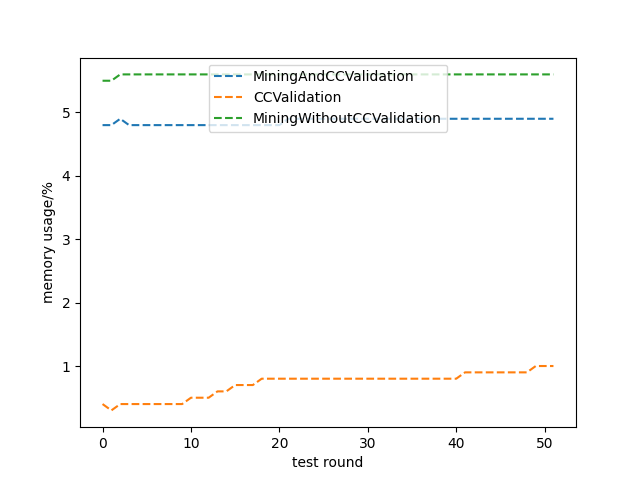}
  \caption{The memory occupation comparison. }
  \label{resourceOccupationMemory}
\end{figure}

\begin{figure}
  \includegraphics[width=3.5in]{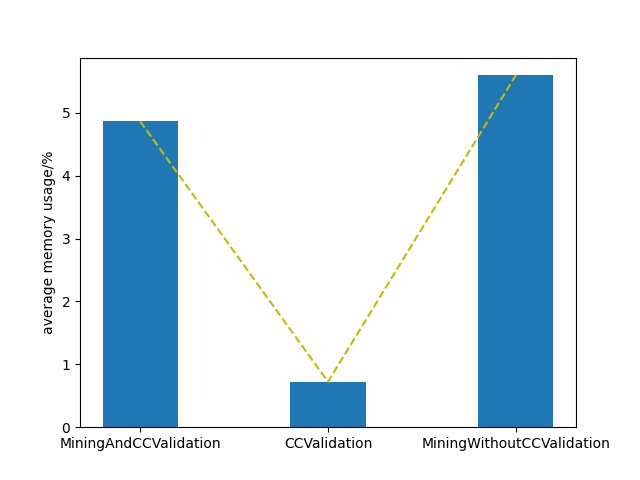}
  \caption{The average memory occupation comparison.}
  \label{resourceOccupationMemoryAverage}
\end{figure}

We collected over 200 data points on resource consumption during the verification process. However, for clarity, we present the results for the first 50 rounds in Figures \ref{resourceOccupationCPU}, \ref{resourceOccupationCPUAverage}, \ref{resourceOccupationMemory}, and \ref{resourceOccupationMemoryAverage}.

The analysis of resource utilization depicted in the figures indicates that the requirements for nodes are within reasonable bounds. Specifically, the average CPU utilization is 104.91\%, 1.26\%, and 97.17\% for "MiningAndCCValidation," "CCValidation," and "MiningWithoutCCValidation," respectively. Cross-chain smart contract validation contributes only a 1.2\% additional CPU load, which is manageable for blockchain nodes.

Regarding memory usage, the average memory utilization is 4.86\%, 0.72\%, and 5.59\% for "MiningAndCCValidation," "CCValidation," and "MiningWithoutCCValidation," respectively. Nodes exclusively performing cross-chain smart contract validation utilize less than 1\% of memory, while nodes involved in both mining and cross-chain validation use less than 5\% of memory, even less than nodes exclusively engaged in mining. This discrepancy may be attributed to random factors, as both threads occupy minimal memory (less than 6\%).

Consequently, considering the CPU and memory usage of cross-chain validation, performing cross-chain smart contract validation for blockchain nodes appears to be feasible from a resource perspective.

\subsubsection{Resource Optimization by Collective Validation}
In this verification scenario, we intend to demonstrate the storage savings achieved through collective validation of smart contracts. Within a local network, one node is designated as the storage node, while other nodes retrieve data for validation from the storage node, eliminating the need to store blockchain data locally.

Three selected nodes, referred to as node 1, node 2, and node 3, are utilized for this purpose. Node 3 is responsible for storing cross-chain validation data, while nodes 1 and 2 request pertinent information from node 3 for cross-chain smart contract validation.

\begin{figure}
  \includegraphics[width=3.5in]{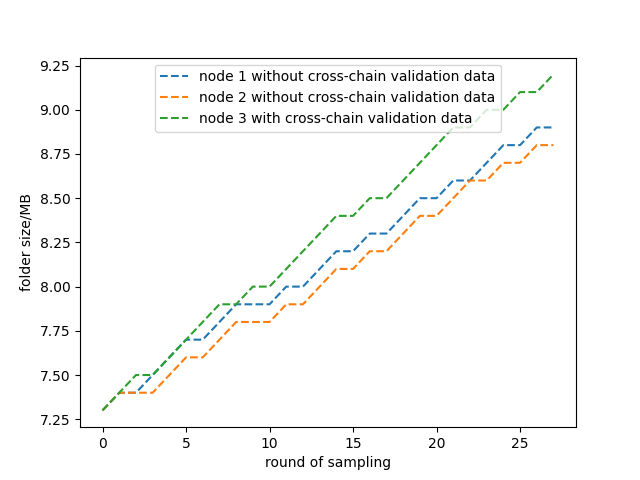}
  \caption{Resources occupation for nodes with shared blockchain data. Three nodes node 1, 2 and 3 shared the blockchain data. Node 3 store the cross-chain validation data for other two nodes.}
  \label{sharedBlockchainData}
\end{figure}

Transactions are sent by a script at one-second intervals in the producer blockchain. Every 30 seconds, the size of the working folder in the consumer blockchain, which contains the blockchain code, logs, and data (local blockchain data and cross-chain validation data), is recorded. The first 30 rounds of sampling are then displayed in Figure \ref{sharedBlockchainData}.

Figure \ref{sharedBlockchainData} depicts the results, showing that the node responsible for storing the cross-chain validation data (node 3) has the largest storage footprint. The average data usage for nodes engaged in blockchain data storage is 8.16MB, 8.08MB, and 8.33MB, respectively, with storage savings exceeding 7\% for one node. As more nodes participate, the storage savings increase, with an average space-saving ratio of $n$. When there are $m$ nodes, this approach can save $n*m$ storage space. Nevertheless, utilizing more nodes to share the same data introduces heightened risks, rendering this approach suitable for small groups. In cases where numerous nodes depend on a single storage node, the failure of the storage node can impact all connected nodes.

\subsection{Impact of Cross-chain Block Segment Length on Cross-chain Smart Contract Validation}
In this section, we examine the influence of the cross-chain block segment length on cross-chain validation, with a focus on two critical aspects: the likelihood of falsifying cross-chain blocks and the issue of competitive rebranching.

\subsubsection{Probability of Falsifying Cross-Chain Blocks} \label{faking_one_block_cheating}
The primary objective is to assess the probability of an adversary node fabricating a block to deceive the system before the synchronization of the next block segment from the producer blockchain. As each segment usually contains a specific number of cross-chain blocks, it takes more than the regular block time to mine a single block. Consequently, there exists a probability for adversary nodes to mine a block with fraudulent data on the consumer blockchain before legitimate blocks are mined.

For our experiments, we use separate processes to represent different blockchain nodes, thereby obviating the need for too many separate hardware devices. These nodes are divided into two groups: Group A represents the producer blockchain with $n$ nodes (as described later), and Group B consists of a single adversary node attempting to mine a fraudulent block in the consumer blockchain. The success or failure of the adversary node's attempt depends on whether it can mine one block before a segment of blocks is mined on the producer blockchain.

We aimed to investigate different numbers ($n$) of nodes on the producer blockchain, including scenarios with up to 2048 nodes, which was challenging to replicate in our lab environment. As the mining time for a single block followed an exponential distribution \cite{fullmer2018analysis}, we were able to simulate this mining time. The average mining time (${averageMiningTime}$) for one block was set to 10 seconds, representing the laboratory's average blockchain mining duration. This parameter was utilized for generating mining times using Python's \textbf{random.expovariate}(1 / ${averageMiningTime}$) function.

Our experimental setup utilized a Lenovo workstation (ThinkStation P350) equipped with 16 CPUs, an 11th Gen Intel(R) Core(TM) i9-11900K processor running at 3.50GHz, and 32 GB of RAM. The operating system used was Ubuntu 22.04.1 LTS.

\begin{figure}[htb]
  \includegraphics[width=3.5in]{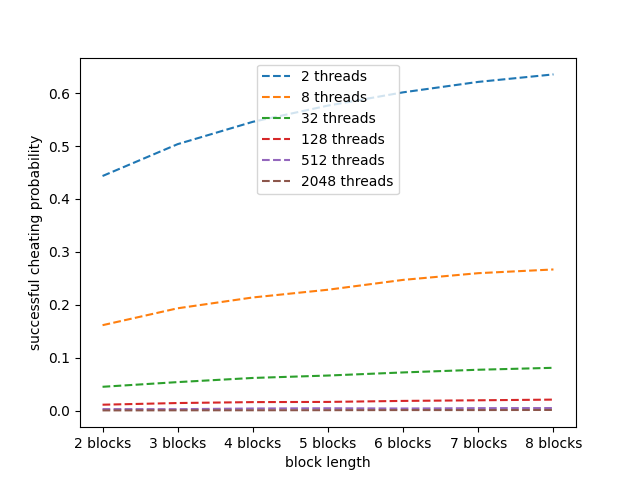}
  \caption{Impact of different lengths on cross-chain block cheat}
  \label{CompareMiningTimeByDifferentBlockListLength}
\end{figure}

To assess the impact of varying cross-chain block segment lengths (\(lengthCrossBlockSegment\)), we considered lengths of 2, 3, 4, 5, 6, 7, and 8. Furthermore, we explored the influence of different producer node counts (\(n\)) in Group A, ranging from 2 to 2048. We conducted 100,000 simulations for each scenario, and the results are shown in Figure \ref{CompareMiningTimeByDifferentBlockListLength}.

Figure \ref{CompareMiningTimeByDifferentBlockListLength} illustrates that with the increase in the length of the cross-chain block segment, the likelihood of successful block cheating (referred to as the successful cheating probability) also increases. For instance, when \(n\) is 2 (indicating two nodes in the producer blockchain for mining), the successful cheating probability ranges from 44.33\% to 63.52\%. However, the incremental gain in  successful cheating probabilities becomes less pronounced as \(n\) increases, primarily due to the already relatively small base probability. When \(n\) reaches 2048, the successful cheating probability is observed to be in the range of 0.06\% to 0.15\%.

Notably, the successful cheating probability changes with different trends. We observed a substantial reduction in the probability of successful cheating as \(n\) increased from 2 to 128, followed by a more gradual decrease from 128 to 2048, where the total success probability remained relatively low. Specifically, when \(n\) increased to 1024, the likelihood of successful block cheating did not exceed 6\%.

\begin{table*}[htb]
  \centering
  \caption{Successful block cheat ratio}
  \label{successfulBlockCheatRatio}
  \begin{tabular}{|c|c|c|c|c|c|c|c|}
  \hline
  \multirow{2}{*}{Nodes} & \multicolumn{7}{c|}{Number of Blocks} \\ \cline{2-8}
   & 2 & 3 & 4 & 5 & 6 & 7 & 8 \\ \hline
  2 nodes & 0.44337 & 0.50355 & 0.54573 & 0.57632 & 0.6014 & 0.62111 & 0.63526 \\ \hline
  8 nodes & 0.16161 & 0.19355 & 0.21398 & 0.22858 & 0.24717 & 0.25985 & 0.26686 \\ \hline
  32 nodes & 0.04526 & 0.05411 & 0.06196 & 0.06652 & 0.07237 & 0.07745 & 0.08112 \\ \hline
  128 nodes & 0.01143 & 0.01458 & 0.01633 & 0.01669 & 0.0186 & 0.01976 & 0.02105 \\ \hline
  512 nodes & 0.00312 & 0.00298 & 0.00426 & 0.0047 & 0.00426 & 0.00491 & 0.0051 \\ \hline
  2048 nodes & 0.00067 & 0.00084 & 0.00094 & 0.00109 & 0.00129 & 0.00129 & 0.00154 \\ \hline
  \end{tabular}
\end{table*}

In summary, the above results revealed several interesting points. We observed that the length of cross-chain block segments has a significant impact on the successful cheating probability, with longer segments leading to higher successful cheat probability. Additionally, as the number of nodes in the producer blockchain increased from 2 to 128, there was a substantial reduction in successful cheating probabilities. However, beyond 128 nodes, the decrease in success rates became more gradual. 

\subsubsection{Impact on Rebranch Probability}
In this section, we focus on investigating the impact of the length of cross-chain block segments on rebranch probabilities. The experimental setup is similar to the previous section, except that in Group B, the adversary node will mine the same number of blocks as that in Group A to trigger a rebranch.

\begin{figure}[htb]
\centering
\includegraphics[width=3.5in]{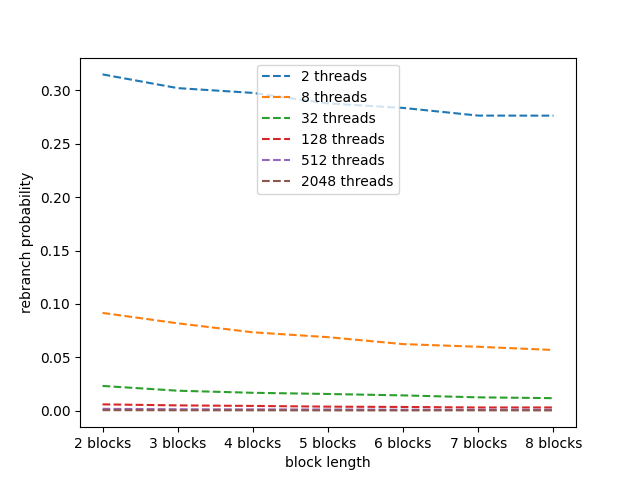}
\caption{Comparison of Rebranch Probabilities}
\label{CompareRebranchProbability}
\end{figure}

The parameters in detail include:
Various lengths of cross-chain block segments (${lengthCrossBlockSegment}$) on rebranch probabilities, ranging from 2 to 8.
Different node counts within the producer blockchain (${n}$) in Group A, spanning from 2 to 2048, across various verification scenarios. Each scenario involves 100,000 simulations, and the results are illustrated in Figure \ref{CompareRebranchProbability}.

Figure \ref{CompareRebranchProbability} illustrates that as the length of the cross-chain block segment increases, the probability of rebranching decreases. For instance, when $n$ is 2, the rebranch probability ranges from 31.50\% to 27.63\%. However, as $n$ increases, the reduction in the rebranch probability becomes less pronounced due to the relatively small base probability. For instance, when $n$ reaches 2048, the rebranch probability decreases from 0.03\% to 0.01\%. Detailed values can be found in Table \ref{rebranch_probabilities}.

\begin{table*}[htb]
    \centering
    \caption{Rebranch Probabilities by different lengthd of cross-chain block segment}
    \label{rebranch_probabilities}
    \begin{tabular}{|c|c|c|c|c|c|c|c|}
    \hline
    \multirow{2}{*}{Nodes} & \multicolumn{7}{c|}{Number of Blocks} \\ \cline{2-8}
     & 2 & 3 & 4 & 5 & 6 & 7 & 8 \\ \hline
    2 nodes & 0.31504 & 0.30219 & 0.2977 & 0.28779 & 0.28364 & 0.27643 & 0.27639 \\ \hline
    8 nodes & 0.09153 & 0.08178 & 0.07334 & 0.06879 & 0.06229 & 0.05978 & 0.05678 \\ \hline
    32 nodes & 0.02308 & 0.01861 & 0.01666 & 0.01555 & 0.01422 & 0.01240 & 0.01162 \\ \hline
    128 nodes & 0.00582 & 0.00485 & 0.00436 & 0.00367 & 0.00338 & 0.00291 & 0.00291 \\ \hline
    512 nodes & 0.00161 & 0.00123 & 0.00111 & 0.00101 & 0.00079 & 0.00073 & 0.00073 \\ \hline
    2048 nodes & 0.00039 & 0.00029 & 0.00025 & 0.00018 & 0.00017 & 0.00025 & 0.00016 \\ \hline
    \end{tabular}
\end{table*}

The findings from Figure \ref{CompareRebranchProbability} and table \ref{rebranch_probabilities} emphasize that longer cross-chain block segments lead to a reduced probability of rebranching. Additionally, while an increase in $n$ results in a lower rebranch probability, the rate of reduction is less significant due to the initially small rebranch probability. These findings contribute to a better understanding of the dynamics of rebranching within blockchain networks.

\section{Conclusion} \label{sec_conclusion}
Conclusion: This paper addresses the critical challenges associated with the validation of cross-chain smart contract results. It introduces a novel method that involves executing the cross-chain smart contract to ensure the integrity of results, effectively mitigating the risks of fraudulent activities during cross-chain propagation. Moreover, the proposed approach emphasizes the significance of adopting a unified view for cross-chain smart contract results through a comprehensive confirmation process, incorporating proof of the chain of blocks and related data from the producer blockchain into the consumer blockchain. The verification results clearly demonstrate the feasibility of conducting cross-chain verification of results at the smart contract level, while also highlighting the manageable resource utilization associated with the proposed validation approach.

Further research can focus on optimizing synchronization protocols to enhance the efficiency and reliability of cross-chain data transmission. Exploring the integration of blockchain technology with emerging fields like Internet of Things (IoT) and decentralized finance (DeFi) can open up new avenues for cross-chain smart contract validation. Additionally, addressing scalability challenges and devising innovative solutions to enable seamless integration among multiple blockchain networks will be pivotal for fostering widespread adoption and interoperability across diverse blockchain ecosystems.

\section*{Acknowledgment}
This work was supported in part by the National Natural Science Foundation of China under Grant No. 61772352; the Science and Technology Project of Sichuan Province under Grant No. 2019YFG0400, 2018GZDZX0031, 2018GZDZX0004, 2017GZDZX0003, 2018JY0182, 19ZDYF1286, 2020YFG0322, and the R\&D Project of Chengdu City under Grant No. (2019-YF05-01790-GX).

\ifCLASSOPTIONcaptionsoff
  \newpage
\fi

\bibliographystyle{IEEEtran}
\bibliography{ref}

%

\begin{IEEEbiography}{Hong Su}
  Hong Su Hong Su received the MS and PhD degrees, in 2006 and 2022, respectively, from Sichuan University, Chengdu, China. He is currently a researcher of Chengdu University of Information Technology Chengdu, China. His research interests include blockchain, cross-chain and smart contract.
\end{IEEEbiography}




\end{document}